# Probing dark excitons in atomically thin semiconductors via near-field coupling to surface plasmon polaritons

You Zhou[1,2*], Giovanni Scuri[2*], Dominik S. Wild[2*], Alexander A. High[1,2*], Alan Dibos[1], Luis A. Jauregui[2], Chi Shu[2], Kristiaan De Greve[1,2], Kateryna Pistunova[2], Andrew Joe[2], Takashi Taniguchi[3], Kenji Watanabe[3], Philip Kim[2†], Mikhail D. Lukin[2†] & Hongkun Park[1,2†]

[1]Department of Chemistry and Chemical Biology and [2]Department of Physics, Harvard University, Cambridge, MA 02138, USA

[3]National Institute for Materials Science, 1-1 Namiki, Tsukuba 305-0044, Japan

*These authors contributed equally to this work.

†To whom correspondence should be addressed: hongkun_park@harvard.edu, lukin@physics.harvard.edu, pkim@physics.harvard.edu

**Transition metal dichalcogenide (TMD) monolayers are direct bandgap semiconductors that feature tightly bound excitons, strong spin-orbit coupling, and spin-valley degrees of freedom[1-5]. Depending on the spin configuration of the electron-hole pairs, intra-valley excitons of TMD monolayers can be either optically bright or dark[6-9]. Dark excitons involve nominally spin-forbidden optical transitions with zero in-plane transition dipole moment[10], making their detection with conventional far-field optical techniques challenging. Here, we introduce a new method for probing the optical properties of two-dimensional (2D) materials via near-field coupling to surface plasmon polaritons (SPPs), which selectively enhances optical transitions with dipole moments normal to the 2D plane. We utilize this**

**method to directly detect dark excitons in monolayer TMDs[8]. When a WSe$_2$ monolayer is placed on top of a single-crystal silver film[11], its emission into near-field-coupled SPPs displays new spectral features whose energies and dipole orientations are consistent with dark neutral and charged excitons. The SPP-based near-field spectroscopy significantly enhances experimental capabilities for probing and manipulating exciton dynamics of atomically thin materials.**

When an optical dipole is in proximity to a metallic substrate, it can emit light into both far field photons and SPPs. Far-field emission can be measured directly via top-down optical microscopy, whereas SPP emission can be detected by converting SPPs into far-field light via engineered out-coupling structures (Fig. 1a). On a single-crystal silver film, our metal of choice due to its low loss[11], SPPs are strongly polarized in the out-of-plane ($z$) direction in the visible frequency range (see Supplementary Information). Consequently, the emission rate into SPPs for an out-of-plane dipole can be as high as 30 times larger than that of an in-plane dipole (Figs. 1b-d, for details of the analysis see Supplementary Information). At the same time, far-field emission of an in-plane dipole is strongly suppressed (Figs. 1b and d) because the in-plane electric field is close to zero near the silver surface. We note that when a point dipole is close to a metal[12], non-radiative recombination due to ohmic loss can be the dominant decay mechanism. Remarkably, for delocalized excitons in quantum wells and 2D materials, quenching of exciton luminescence by ohmic loss is significantly reduced, even when they are placed 10 nm above a silver surface ( see [13] and Supplementary Fig. 1 and discussion). Combined together, the net effect of a nearby silver surface is significantly enhanced (suppressed) emission of an out-of-plane (in-plane) dipole into SPPs (far field).

In our experiment, we employ exfoliated monolayers of WSe$_2$ or MoSe$_2$ encapsulated between thin, insulating hexagonal boron nitride (hBN) flakes[14]. The photoluminescence (PL) spectra of these heterostructures exhibit narrow excitonic features (linewidths of 2 ~ 4 meV) at 4 K. A TMD monolayer encapsulated by hBN is placed on atomically smooth single-crystal silver[11]. In our devices, silver plays three separate roles: it acts as a mirror for far-field detection, supports SPPs that can be coupled to TMD excitons, and serves as a gate electrode that can change the electrostatic potential of a TMD monolayer (Fig. 1a). The spacing between the monolayer TMD and the silver surface is determined by the bottom hBN thickness, and can easily be controlled by varying hBN thickness. In our devices, the typical spacing is on the order of ten nanometers.

Excitons are created using off-resonant 660-nm laser excitation, and the PL spectra are collected either from the same spot excited by the laser via far-field imaging (FF-PL), or from scattered SPPs at the out-coupling structures (SPP-PL, inset of Fig. 1e). At 4 K, the FF-PL spectrum of WSe$_2$ display several commonly observed spectral features, assigned previously to neutral exciton emission ($X_0$), charged exciton emission ($X_T$), and the lower-energy emission (L1) often attributed to defects[15]. Both $X_0$ and $X_T$ are known to have only in-plane transition dipole moments[16]. The SPP-PL spectrum, however, exhibits an additional large peak ($X_D$) at 731 nm that is absent in the far-field spectrum, with narrower linewidth (~2 meV) compared with other spectral features. In contrast, neither FF-PL nor SPP-PL spectra of monolayer MoSe$_2$ exhibit the feature corresponding to $X_D$ in WSe$_2$ SPP-PL (see Supplementary Fig. 2). These observations suggest that the WSe$_2$-specific $X_D$ feature is associated with an out-of-plane transition dipole moment that preferentially couples to SPPs.

To further explore the nature of excitonic species in monolayer TMDs, we compare the FF- and SPP-PL spectra for monolayer WSe$_2$ and MoSe$_2$ as we tune the carrier density via a gate voltage. It is well known that neutral exciton emission ($X_0$) of TMD monolayers only exists inside the band gap[17]. Once electrons or holes are added to the conduction band (CB) or valence band (VB) of TMD monolayers via electrostatic doping by a gate voltage, the emission from $X_0$ vanishes, and the PL spectra are dominated by the charged exciton emission, $X_T$. In Figs. 2a-d and in Supplementary Figs. 2 and 3, such trends are evidenced by the simultaneous disappearance of $X_0$ and the onset of in-plane conduction outside the bandgap region (-2.5 V < $V_g$ < 0 V)[18]. Gate-dependent SPP-PL spectra of a WSe$_2$ monolayer in Fig. 2b shows that emission from $X_D$ exists only within the same voltage range as the bright neutral exciton $X_0$.

Outside the bandgap region, additional optical transitions with an out-of-plane dipole orientation emerge. Figures 3a-b illustrate FF- and SPP-PL spectra of WSe$_2$ at different gate voltages. We normalize both FF and SPP-PL spectra using the intensity of a charged exciton peak $X_T$ because it is known to involve a purely in-plane transition dipole moment[16]. The ratio of SPP-PL intensity to the FF-PL intensity after the normalization provides a direct measure of the orientation of the transition dipole for each luminescent species: the unity ratio represents a purely in-plane dipole, while a value larger than one indicates that the transition dipole has some out-of-plane components. Based on our theoretical calculations presented in Fig. 1d and Supplementary Fig. 4, an optical transition with a purely out-of-plane transition dipole should have a normalized coupling ratio of 7 in our device geometry. The experimental results for $X_D$ yield a value of 16: this discrepancy between theory and experiment is likely due to small, yet non-negligible absorption of SPPs by charged excitons as they propagate through the WSe$_2$, which increases the apparent coupling ratio of $X_D$ after normalization (see Supplementary Fig.

5). Indeed, when SPPs propagate through a minimal distance within WSe$_2$ (Supplementary Fig. 6), the normalized coupling ratio determined by experiment is close to 7, in good agreement with the theoretical calculations. Figure 3b shows that the value of the normalized coupling ratio in WSe$_2$ is large not only for the neutral exciton peak X$_D$ at 731 nm, but also for the L1 peak at 740 nm. Specifically, in the band gap region (-2.5 V < $V_g$ < 0 V), the normalized coupling ratio of L1 is ~ 3, but increases significantly and reaches a value of ~16 just when the Fermi level reaches the VB maximum ($V_g$ ~ -2.5 V) and the CB minimum ($V_g$ ~ 0 V). A plot of normalized coupling ratios for MoSe$_2$ does not show comparable feature (Supplementary Figure 7).

The origin of optical transitions with an out-of-plane dipole moment (X$_D$ and L$_1$) in WSe$_2$ can be understood by examining the band structure of monolayer TMDs. At the CB minimum and VB maximum of monolayer TMDs, *i.e.* the $K$ and $K'$ points in reciprocal space, the electronic wave functions can be characterized by their transformation properties under a 120° rotation ($R$) about the surface normal ($z$ axis), and a reflection ($P$) in the TMD plane ($xy$ plane, Fig. 4a). Each orbital wave function carries a magnetic quantum number $m$ (only $m$ modulo 3 matters because of the reduced rotational symmetry) and a parity $p$ (odd: $p = -1$; even: $p = 1$). Selection rules dictate that light circularly polarized in the $xy$ plane ($\sigma_\pm$) can only induce transitions between states with the same parity and magnetic quantum numbers differing by one. By contrast, light polarized linearly along the $z$ axis couples states with opposite parity and identical magnetic quantum numbers.

Monolayer TMDs exhibit strong spin-orbit coupling (SOC) proportional to **L**·**S**, where **L** and **S** are the orbital and spin angular momentum, respectively. The spin-orbit interaction leads to a large spin splitting in the VBs (on the order of several hundred meV[19-21]) and a much smaller splitting in the CBs. The latter is because the spin degeneracy in the CB is only lifted to second order[6,7,9] by the SOC term $L_+S_- + L_-S_+$, which mixes different orbital and spin wave functions. Notably, the splitting is of opposite sign for WSe$_2$ and MoSe$_2$[6,7,9] (the band orderings for WSe$_2$ and MoSe$_2$ are shown in Figs. 4b and c). Owing to the SOC mixing term, spin is no longer an exact quantum number, and nominally spin-forbidden transitions, associated with dark exciton states, are in principle possible, albeit weakly. We note that the spin directions in Figs. 4b and 4c only denote the dominant spin direction. Two out of the four possible transitions at the K point are between bands with the same dominant spin directions and are therefore optically "bright". For the remaining two transitions that are nominally "dark", the transition dipole moment should be along the z direction because SOC only couples states with opposite parity. Including magnetic quantum numbers, group theoretic considerations dictate that only the transition connecting to the upper valence band is allowed with z-polarized light[10]. The strength of the out-of-plane dipole moment is predicted to be weak, approximately a hundred times smaller than the in-plane bright exciton dipole moment[10]. However, because the energy scale associated with the sample temperature (~0.4 meV at 4 K) is much smaller than the bright/dark exciton energy splitting of ~40 meV, the exciton population at thermal equilibrium is almost entirely in the dark exciton state[8], and the dark-exciton emission should be visible as long as the z-polarized light can be collected, as is the case of SPP-PL.

These band-structure and SOC considerations strongly suggest the $X_D$ feature we observe in SPP-PL of WSe$_2$ (Figs. 2b and 3), with its z-polarized dipole orientation, corresponds to the dark exciton described in Fig. 4b. In contrast, in MoSe$_2$ SPP-PL (Fig. 2d), the corresponding feature is absent (Supplementary Figs. 2 and 7) because the dark exciton has a higher energy than the bright exciton[6-9,22] and thus should not be observed in our experiment. We note that the $X_D$ feature in WSe$_2$ lies 42 meV below the bright exciton transition ($X_0$), similar to recently observed dark excitons[23,24] under the application of a large (up to 30 Tesla) in-plane magnetic field. The small difference in bright/dark exciton energy splitting (42 meV in our case vs 47 meV in magnetic-field studies[23,24]) likely originates from different experimental conditions that can modify the dielectric environment and the band structure of $WSe_2$[25-27]: hBN encapsulated $WSe_2$ on top of silver in our measurements vs unpassivated $WSe_2$ on $SiO_2$[23,24].

The emergence of L1, a *z*-polarized transition outside the bandgap region, may indicate the formation of dark charged excitons. Once the Fermi level reaches the band edge of either the CB or VB, features corresponding to the bright and dark neutral excitons disappear, suggesting that these neutral excitons bind with free carriers to form charged excitons (trions[28,29] or polarons[30]). The peak L1 may thus originate from charged dark excitons that form when neutral dark excitons bind with free electrons and holes. It should be noted that similar suggestions have been made using the observation of L1 brightening under an in-plane magnetic field[23]. The band structure of $WSe_2$ dictates that a dark charged exciton should form through the binding of a dark neutral exciton with an electron/hole in the other valley. The bright charged excitons are formed, on the other hand, by the binding of a bright neutral exciton with an electron/hole from the same valley. Figure 3b also shows significant brightening of the L1 feature as the Fermi level reaches the VB maximum ($V_g \sim -2.5$ V) and the CB minimum ($V_g \sim 0$ V): this observation may suggest the reorientation of a transition dipole moment at those points or alternatively a change in the relative population of states with in-plane versus out-of-plane transition dipole moments. The exact reasons for the observation is not understood at the moment, and requires further experimental and theoretical analyses.

We have demonstrated a new method for SPP-assisted near-field spectroscopy that enables measurement of dipole orientation of 2D excitons. We use this method to observe nominally spin-forbidden dark excitonic states in monolayer WSe$_2$ with out-of-plane dipole orientation, thus directly probing the SOC physics in TMD monolayers. Our method opens up many new avenues for studying and controlling fundamental properties of atomically thin materials. Specifically, this new spectroscopy method should enable the detail studies of indirect excitons[31] and the interaction between out-of-plane excitons and phonons in van der Waals heterostructures[32]. In addition, the method may also enable the new approaches to strong light-matter coupling involving dark excitons. Due to the negligible coupling to free-space electromagnetic modes, the near field coupling to tightly confined nanophotonic and plasmonic modes should play the dominant role in radiative interactions of dark excitons under a wide variety of experimental conditions. These can be used to explore novel types of active metasurfaces, spatial modulation of exciton lifetimes as well as efficient schemes for chiral photonics[11] and quantum optics[33,34]. Finally, the dark excitons can have long lifetime[8,10], which may be employed to study transport and electrostatic manipulation of excitonic matter in 2D systems[30,35,36].

**Acknowledgments**

We acknowledge support from the DoD Vannevar Busch Faculty Fellowship (H.P., M.D.L.), AFOSR MURI, the Gordon and Betty Moore Foundation, NSF, CUA, and ARO MURI. We carried out all film deposition and device fabrication at the Harvard Center for Nanoscale Systems.

## Author contributions

H.P., P.K., M.D., A.H., Y.Z., A.D., and L.J. conceived the study, and Y.Z., G.S., A.H., A.D., L.J., K.P. & A.J., developed the fabrication procedure. Y.Z, G.S, A.H, A.D, C.S & K.d.G. performed experiments, and T.T. and K.W. performed hBN growth. D.W. and A.H. performed computational analyses and simulations, and D.W., M.L. and H.P. contributed to theoretical descriptions. Y.Z., G.S., D.W., A.H., P.K., M.L. and H.P. wrote the manuscript, with extensive input from all authors. The authors declare no competing financial interests.

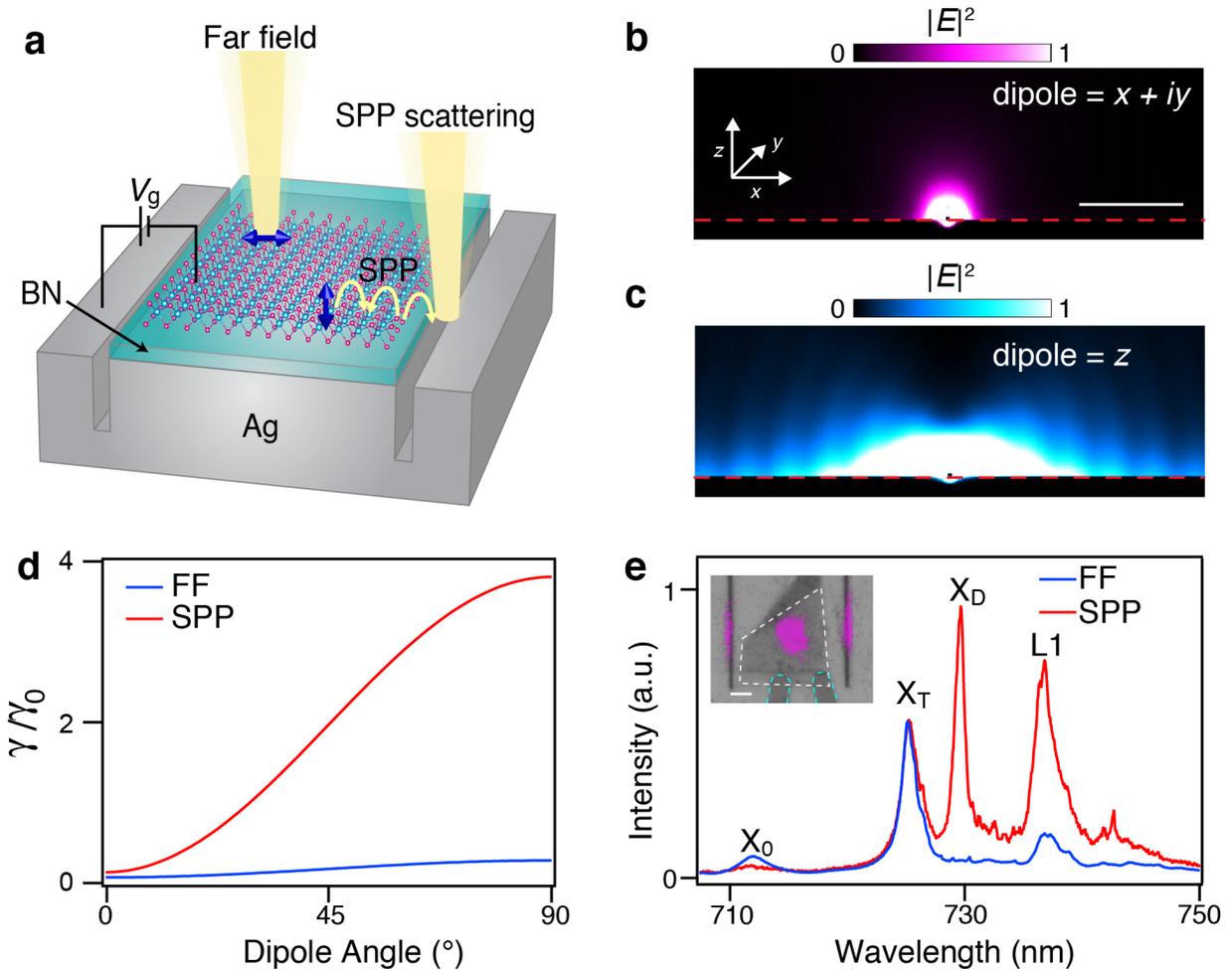

**Fig. 1. Probing out-of-plane electric dipole transitions in 2D materials via near-field coupling to SPPs.** (**a**) Schematic of the experimental setup. A TMD-based van der Waals heterostructure is placed on top of single-crystal silver. Optical transitions can couple to both near-field SPPs and far field. The coupling to SPPs can be detected by scattering the SPPs into the far field using nanoscale trenches etched into silver. (**b**) and (**c**) Finite-difference time-domain (FDTD) simulation of the electric-field intensity radiated by (b) an in-plane point dipole emitter and (c) an out-of-plane point dipole emitter. The emitter is positioned 10 nm above the silver surface. Scale bar: 1μm. (**d**) Emission rate into the far field (blue) and SPPs (red) as a function of the angle $\theta$ between the transition dipole and the 2D plane (0° is defined as in the 2D plane). $\gamma_0$ is the decay rate of the exciton population in vacuum at $T = 4$ K. In the analysis, the dipole emitter is 8 nm above a silver film, with 8 nm (5 nm) thick hBN below (above) the dipole to match the heterostructure geometry in the experiment. The emission wavelength is 720 nm. Far-field emission is defined as emission into an objective lens above the sample with numerical aperture of 0.75, to match the experimental measurement geometry. **(e)** PL spectra of WSe$_2$ collected from far field (blue) and from scattered SPPs (red) at $T = 4$K and at zero applied gate voltage. Inset: PL intensity map with a fixed incident excitation beam overlaid with an optical image of a monolayer WSe$_2$ device on silver. The white-dashed (cyan-dashed) line indicates the region of WSe$_2$ (electrical contacts). Scale bar: 1 μm.

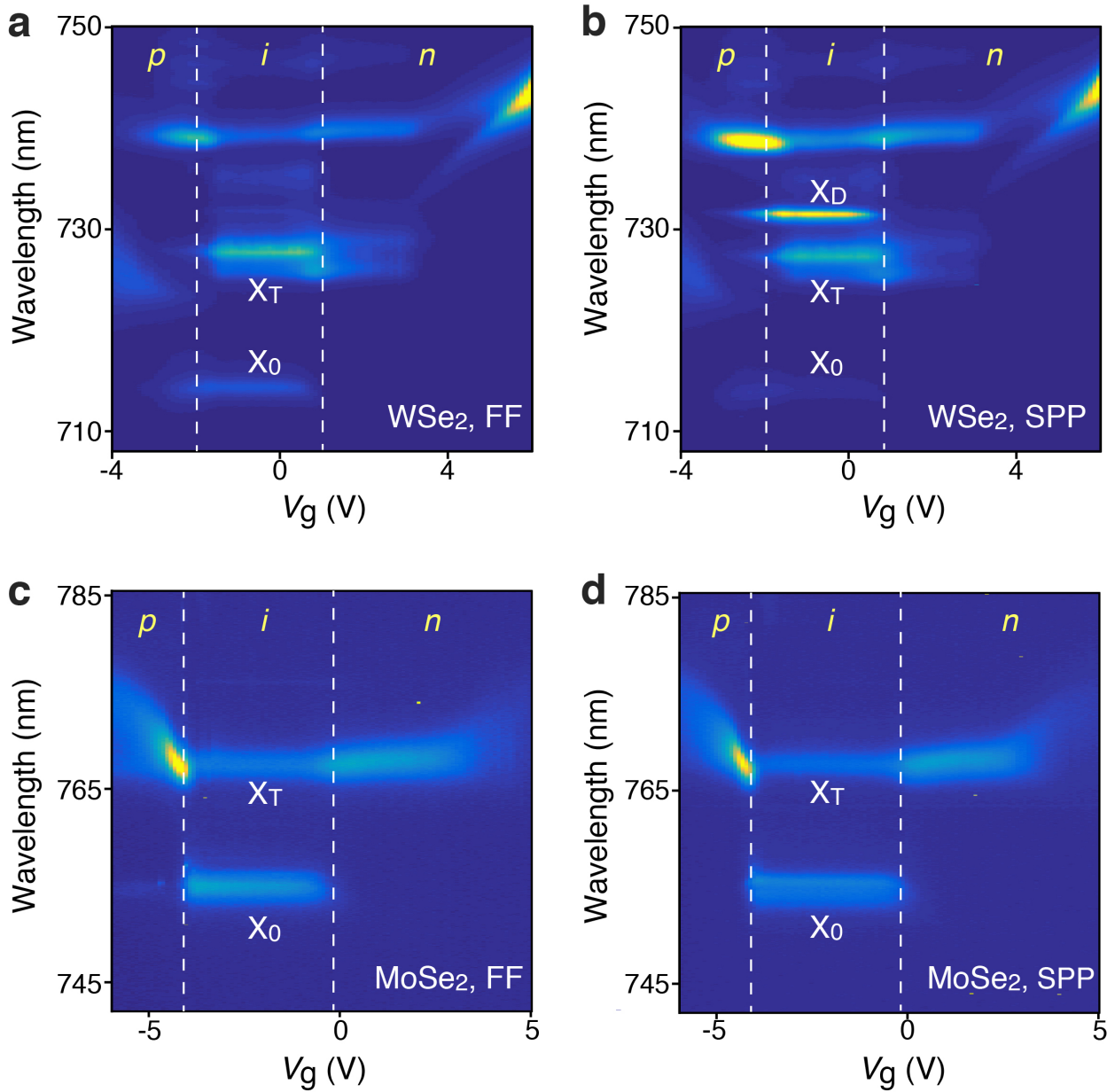

**Fig. 2. Gate dependent FF- and SPP-PL spectra in $WSe_2$ and $MoSe_2$.** **(a)** and **(b)** Gate-voltage ($V_g$) dependent **(a)** FF- and **(b)** SPP-PL spectra of $WSe_2$ and **(c)** FF- and **(d)** SPP-PL spectra of $MoSe_2$. The intensity is plotted in the linear scale from low (dark blue) to high (yellow). The BN gate dielectric thickness is 8 nm for the $WSe_2$ sample and 10 nm for the $MoSe_2$

sample. The PL intensity of (**a**) is normalized with respect to that of (**b**) (and (**c**) with respect to (**d**)) so that the intensity of the charge exciton $X_T$ at zero gate voltage is equal. The dashed lines in the figures correspond to the boundaries where $X_0$ exists in FF-PL, which correlate well with the gate voltages that mark the onset of in-plane conduction. Negative (positive) gate voltages correspond to hole (electron) doping.

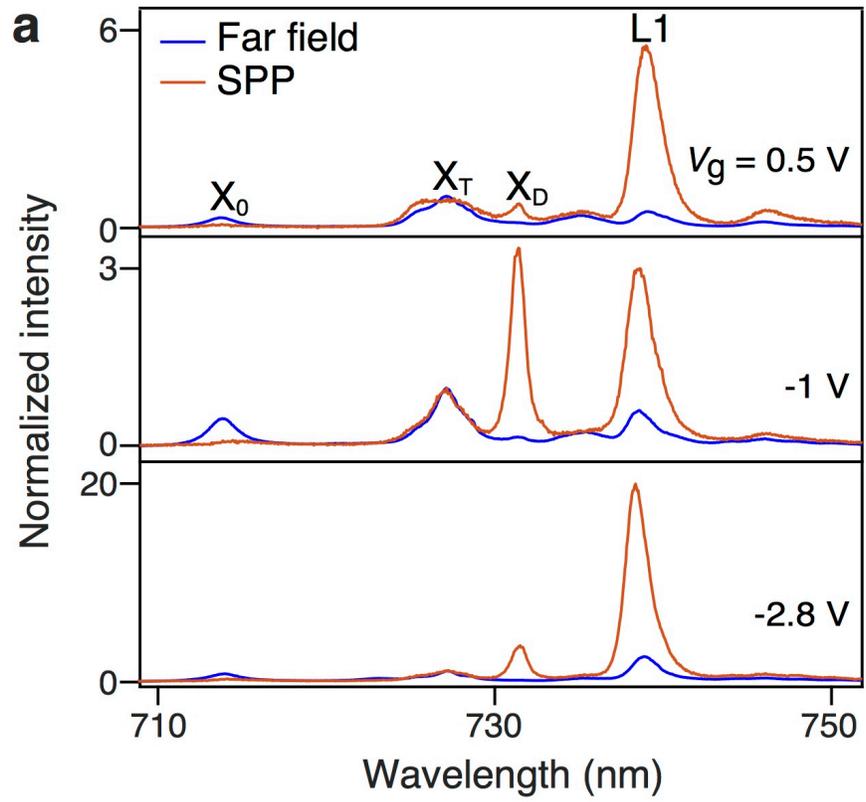

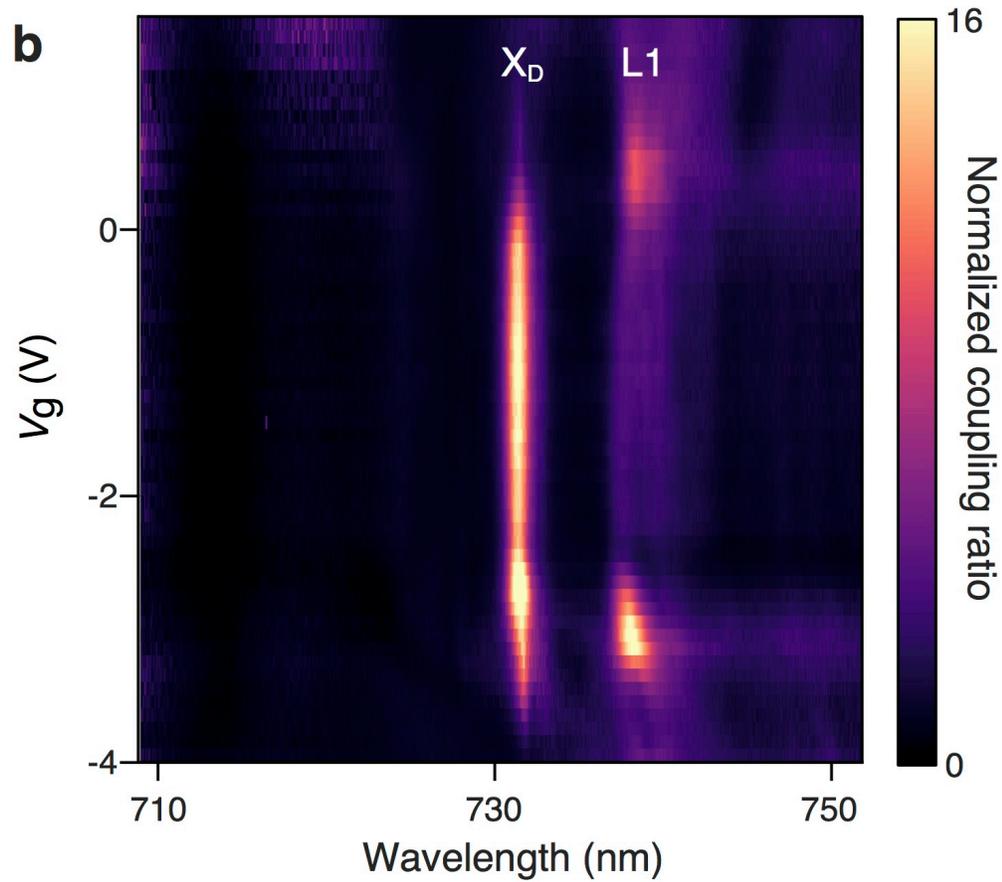

**Fig. 3. Gate-dependent enhancement of charged dark exciton state. (a)** FF- and SPP-PL spectra of WSe$_2$ normalized to the $X_T$ intensity at different gate voltages. **(b)** The normalized coupling ratio, defined by the SPP-PL intensity divided by FF-PL intensity both normalized by the bright trion intensity, as a function of gate voltage.

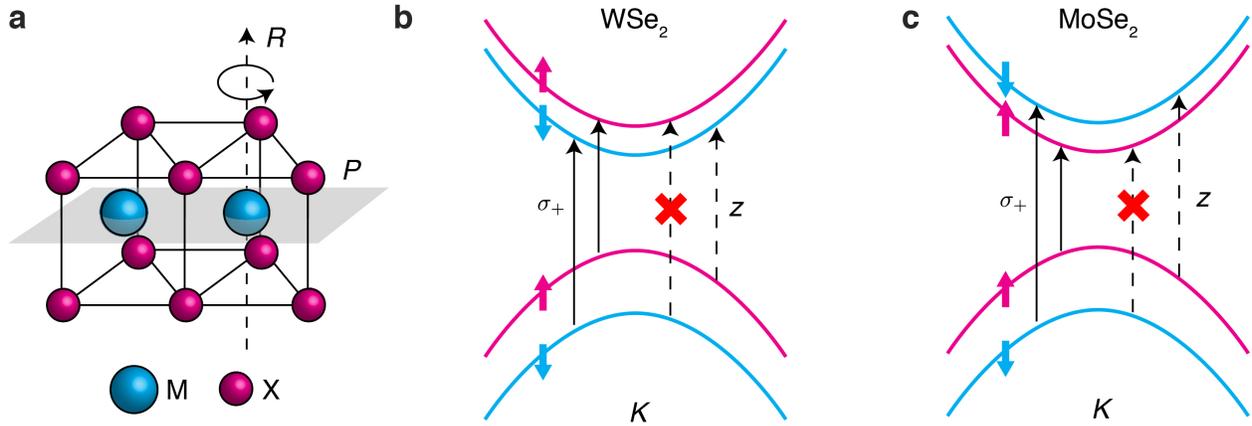

**Fig. 4. Band structure and optical transitions in $WSe_2$ and $MoSe_2$ in the presence of SOC.** **(a)** Crystal structure of TMDs ($MX_2$) and two of its symmetries (3-fold rotation axis $R$ and reflection plane $P$). **(b)** and **(c)** Allowed transitions at the $K$ point for **(b)** $WSe_2$ and **(c)** $MoSe_2$. The colored arrows indicate the dominant direction of electronic spin. Due to spin-orbit coupling (SOC), spin is not an exact quantum number, and the spin-flip transition between the upper valence band and the lower conduction band is weakly allowed. The transition between the lower valence band and the upper conduction band is strictly dipole forbidden.

**Methods:**

Single-crystal silver films (250-nm thick) were grown via sputtering on the (0001) face of 4H-SiC substrates at 350 ºC with a buffer layer of Ti (90-nm thick) that was grown at 500 ºC (Details of the growth will be reported elsewhere). Typical silver films show a RMS roughness less than 1 nm and exhibit long propagation length for SPPs[11]. Trenches that served as in- and out-coupling structures for SPPs were made using electron-beam lithography and Argon ion etching. Thin layers of hBN and monolayers of $WSe_2$ and $MoSe_2$ were mechanically exfoliated from bulk crystals onto Si wafers coated with 285-nm thick $SiO_2$. Monolayer TMDs were then identified under an optical microscope and verified via PL measurements. The thicknesses of hBN layers were measured by atomic force microscopy. The hBN/TMD/hBN heterostructures were then assembled and transferred by a dry transfer method[37] onto the patterned single-crystal silver substrates. Afterwards, thin layers of SiN (160 nm; via plasma-enhanced chemical vapor deposition at room temperature) and alumina (30 nm; via atomic layer deposition at 90 ºC) were deposited, with the pattern defined by electron-beam lithography. This served as an insulating barrier between the silver substrate and the electrical contacts to the TMDs. Finally electrical contacts were deposited by electron-beam evaporation of Cr (10 nm)/Au (90 nm) followed by sputtering of another 40 nm of Au, with the dimensions defined by electron-beam lithography. Optical measurements were carried out in a home-built confocal microscope using an objective with a numerical aperture of 0.75 in a 4 K cryostat from Montana Instruments. Two galvo mirrors were used to scan the excitation and collection spot independently on the sample.